# Foot-print of Claudin and Occludin Transcriptome in Colorectal Cancer

**Authors:** Maryam Ghoojaei[1], Reza Shirkoohi[2], Mojtaba Saffari[3], Amirnader Emamirazavi[4], Mehrdad Hashemi[5].

1. Department of Genetics, Tehran Medical Sciences Branch, Islamic Azad University, Tehran, Iran. maryam.ghoojaei@gmail.com
2. Cancer Biology Research Center, Cancer Institute of Iran, Tehran University of Medical Sciences, Tehran, Iran.
3. Cancer Biology Research Center, Cancer Institute of Iran, Tehran University of Medical Sciences, Tehran, Iran.
4. Iran National Tumor Bank, Cancer Institute of Iran, Tehran University of Medical Sciences, Tehran, Iran.
5. Department of Genetics, Tehran Medical Sciences Branch, Islamic Azad University, Tehran, Iran.

## Abstract

**Background and Purpose:** Colorectal cancer as a leading cause of mortality worldwide, can be regarded as a relatively common and fatal disease with increasing incidence over recent years. Colorectal cancer is characterized by uncontrolled growth of abnormal cells occurring in different parts of the colon. About 90% of deaths associated with cancers occur due to metastasis which overcomes the body's cellular connection, including tight junctions. Claudin and occludin are integral membrane proteins found in tight junctions. The aim of the present study was to investigate the expression level of claudin and occludin in human colorectal cancer.

**Method:** In this study, 38 colorectal cancer patients who referred to Cancer Institute of Imam Khomeini Hospital in Tehran, Iran were studied after obtaining the informed consent. First, quantitative extraction of RNA was performed, then the expression levels of claudin and occludin genes were examined by reverse transcription, PCR and Real-time PCR.

**Findings:** The expression levels of both claudin and occludin genes in cases with higher stage and grade of disease, in the state of metastasis were more than those of the control samples.

**Conclusion:** The increased expression level of the mentioned genes can be considered as an influential factor in turning the normal healthy tissues into cancerous cells.

**Keywords:** Colorectal cancer, Colon, Large intestine, Metastasis, Gene expression, Claudin, Occludin**.**

**Introduction**

Colorectal cancer is a fatal and relatively common disease in Iran (5000 new cases per year) [1]. In 2015, the number of people suffering from the mentioned cancer across both sexes and all ages has been estimated to be 7,937 [2]. Following the cancer diagnosis, the staging process is used to determine whether the cancer has remained within the intestine or has spread to other parts of the body. Gaining information about the stage of cancer is of great significance to choose the most effective treatment method [3]. As metastasis causes about 90% of cancer deaths, its early detection is helpful to offer the appropriate treatment interventions and consequently control the disease [4]. Metastasis is a complex process that involves the following steps: dissociation of tumor cells from primary tumor, invasion, migration, entrance into blood vessels, survival in the blood vessels, entrance to the parenchymal tissue of target organ, and finally formation of a colony in the secondary location [5]. The metastasis results from various changes in tumor cells and molecular system of the body. The state of epithelial-mesenchymal transition can be regarded as one of the mentioned changes. Recent studies indicate that epithelial mesenchymal transition (EMT) is a stem cell mechanism which increases cell migration and loss of cell polarity during formation and differentiation of stem cells. In cancer, EMT plays different roles in preserving the cancer, inducing metastasis, and prolonging the life of cancer cells over their spread from the initial cancer site to the ultimate metastasis site [6]. In other words, EMT is a phenomenon associated with a chain of events that begins with changes in the cell-to-cell connection and cytoskeleton. This phenomenon can lead to changes of cells in the extracellular matrix and release of epithelial cells from the peripheral tissue [7]. Tight junctions are tough intercellular adhesions with a complex molecular structure including adaptor proteins like zonula occludens and integral membrane proteins like claudin and occludin [8]. Occludin consists of four transmembrane domains, three cytoplasmic domains, and two extracellular loops [9]. Claudins have four transmembrane domains and two extracellular loops through which connections are provided to the adjacent cells, (i.e., claudins) [10]. Considering the increasing frequency of colorectal cancer, the present study aimed at addressing the correlation of EMT phenomenon with metastasis, examining the changes of claudin and occludin occurring in this phenomenon, investigating the expression level of these two genes in the patient's samples, and attending to its relation with the morphological and clinical findings to identify biomarkers indicating invasive cells, which are of clinical value.

**Materials and Method**

The present case-series study tracked the patients who referred to the Cancer Institute of Imam Khomeini Hospital, Tehran, Iran. After obtaining informed consent, 38 specimens from colorectal cancer regardless of the stage of disease were randomly selected from National Tumor Bank of Iran, Cancer institute of Imam Khomeini Hospital Complex, Tehran University of Medical Sciences, Tehran, Iran. The samples were previously examined by pathologists, and TNM staging system was applied to assess the stage of cancer. T indicates tumor size, N expresses lymph node involvement, and M means metastasis.

The RNA was extracted using easy-BLUE RNA extraction kit (iNtRON Bio-technology, Seoul, Korea). The optical density (OD) and absorption of RNA were measured at a wavelength of 260 nm by a nanodrop machine (Spectrophotometer, 2000, Thermo Fisher Scientific Inc., Wilmington, USA). On agarose gel electrophoresis, RNA revealed its good quality.

The cDNA was then prepared using the Random Hexamer primer. As Table 2 indicates, real-time PCR (Rotor-Gene Q real-time PCR cycler, QIAGEN Inc., Valencia CA, USA) was performed to measure the expression of the desired genes using the designed primers (Table 1). To evaluate any changes in the gene expression and to make comparisons, the relative expression software tool (REST, QIAGENE Inc., Valencia) was used. Finally, the statistical results of the study were obtained by calculating $2^{-\Delta\Delta CT}$ and t-test was used to measure the *p*-value. The results were considered to be statistically significant at $p \leq 0.05$.

**Table 1. Primers used in PCR reaction**

| Gene | FW | RWD |
|---|---|---|
| Claudin | 5´-TCGATACAATGGCACAGTGG-3´ | 5´-CAATCCCGCTATTGTGGTTT-3´ |
| Occludin | 5´-TCCAATGGCAAAGTGAATGA-3´ | 5´-GCAGGTGCTCTTTTGAAGG-3´ |
| GAPDH | 5´-TCACCAGGGCTGCTTTTAAC-3´ | 5´-GACAAGCTTCCCGTTCTCAG-3´ |

**Table 2. qPCR reaction temperature cycles**

| Cycle step | Temp. | Time | Cycles |
|---|---|---|---|
| Initial denaturation | 95˚c | 15 min | 1 |
| Denaturation | 95˚c | 15s | 40 |
| Annealing | 55-60˚c | 60s | |
| Elongation | 72˚c | 20s | |

**Results**

Histopathological characteristics of patients were received from the National Tumor Bank of Iran. The average age of patients was 55 years, with the maximum and minimum age of 79 and 16, respectively. Patients' clinical characteristics, including pathology, and tumor size are presented in Table 3.

**Table 3. Patients' profile**

| Variables | No. (%) |
|---|---|
| Age | |
| <40 | 6 (%15.78) |
| ≥40 | 32 (%84.22) |
| Size | |
| <3 | 8 (%21.05) |
| 3≤tumor size≤5 | 16 (%42.10) |
| >5 | 14 (%36.85) |
| Grade | |
| I | 11 (%28.94) |
| II | 18 (%47.36) |
| III | 6 (%15.78) |
| IV | 1 (%2.66) |
| X | 2 (%5.26) |
| Stage | |
| I | 3 (%7.89) |
| II | 16 (%42.11) |
| III | 16 (%42.11) |
| IV | 3 (%7.89) |
| Metastasis | |
| M0 | 29 (%76.33) |
| M1 | 6 (%15.78) |
| MX | 3 (%7.89) |

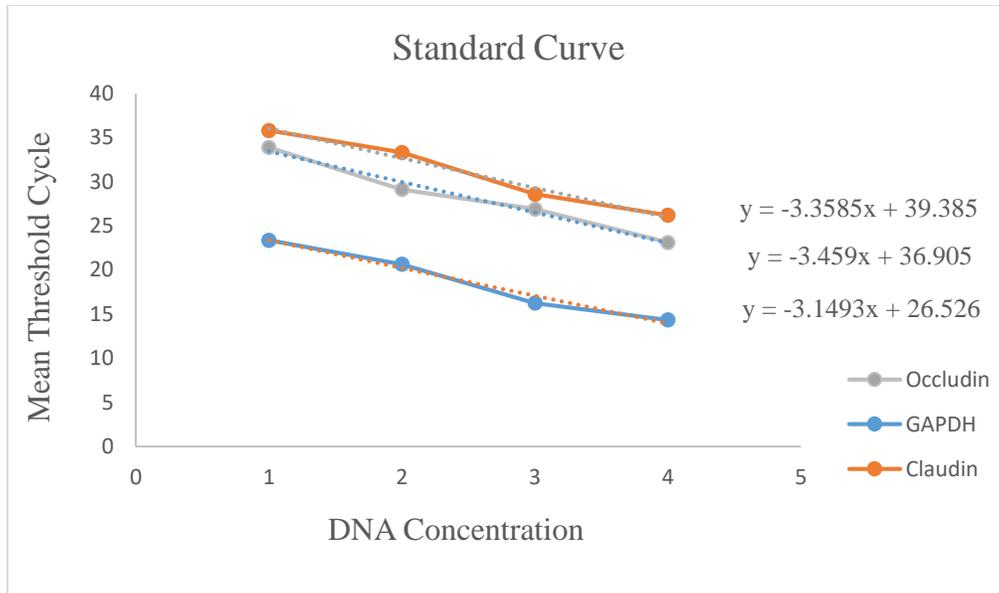

**Figure 1. Standard curves for claudin, occludin and GAPDH**

As shown in Figure.1 the slope of standard curves for claudin, occludin and GAPDH were respectively determined -3.3585, -3.459 and -3.1493. PCR efficiency (E) was calculated 98.516% for claudin, 94.581% for occludin and 107.75% for GAPDH gene.

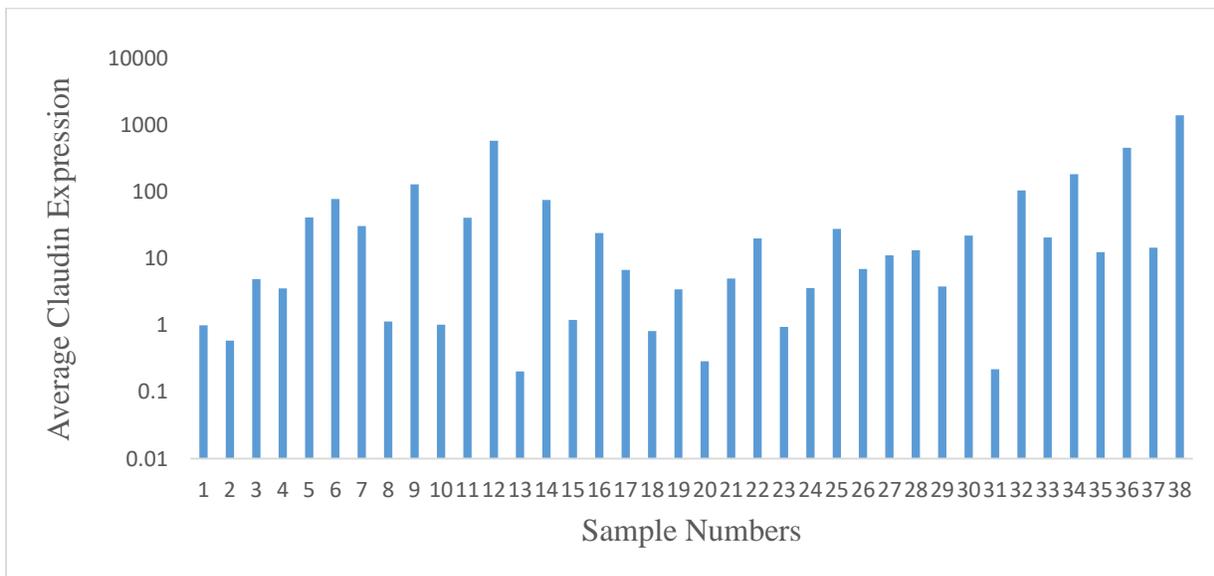

**Figure 2. Changes in claudin gene expression**

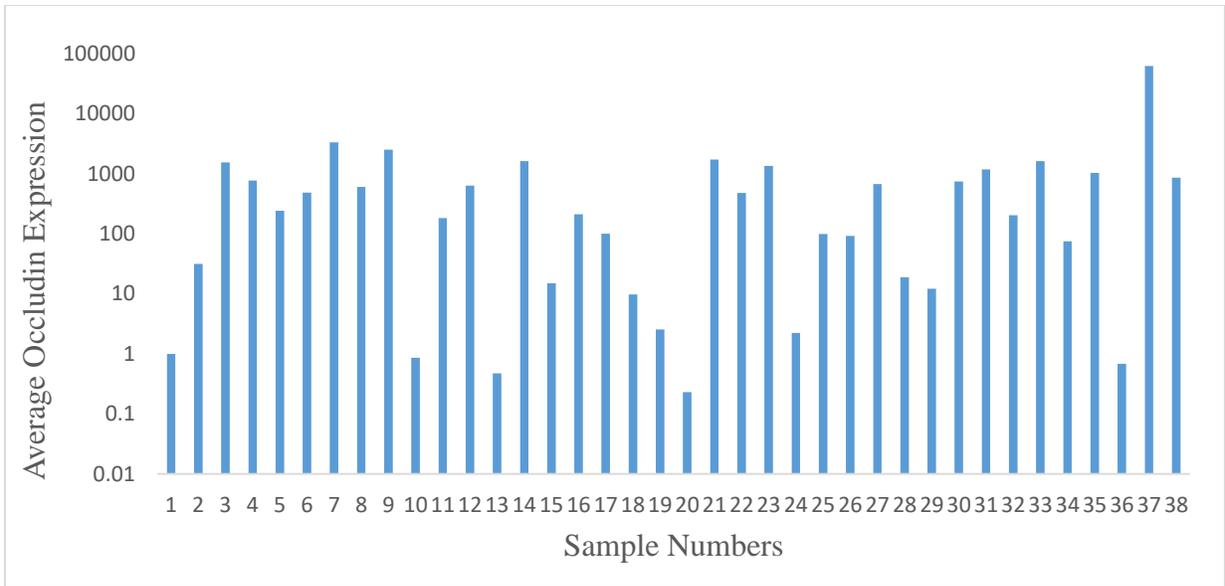

**Figure 3. Changes in occludin gene expression**

As indicated in Figures 2 and 3, the expression levels of claudin and occludin genes increased in most specimens.

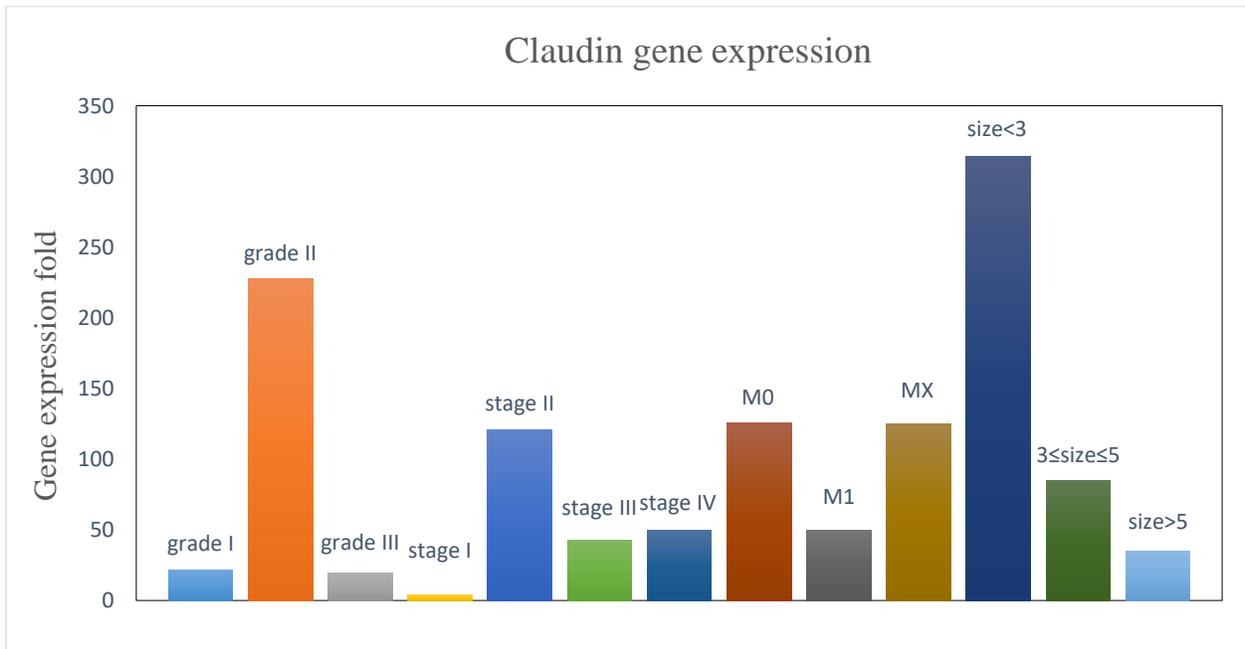

**Figure 4. Average claudin expression according to grade, stage, metastasis and tumor size**

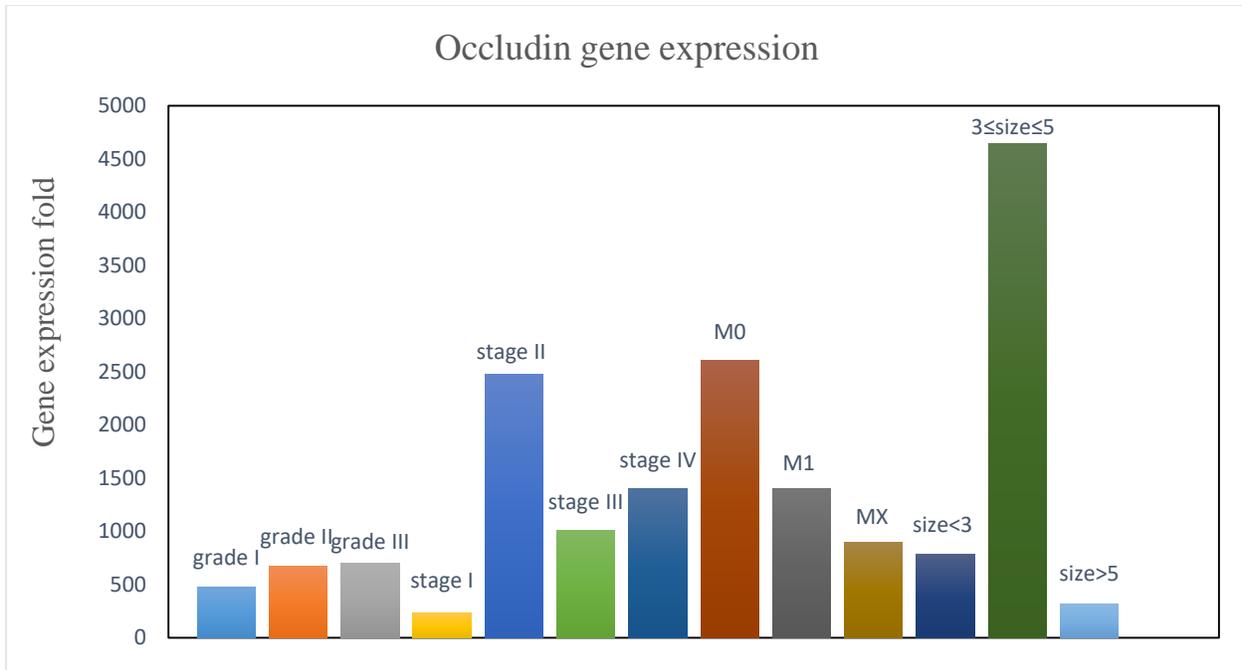

**Figure 5. Average occludin expression according to grade, stage, metastasis and tumor size**

With respect to claudin and occludin expression levels, considering the grade of disease, the majority of samples had higher gene expression levels than the normal sample across all grades. According to Figure 4, the average claudin expression level of tumor samples in grade II is about 10 times more than the other grades, and there is no direct correlation between the grade progression and gene expression level. The average increase of claudin expression level was significant in grades II and III ($P = 0.0377$ and $P = 0.027428$, respectively).

Figure 5 reveals that the average of occludin expression level also increased in parallel with grade progression. The expression level of tumor samples in grade III is about 1.5 times more than the other grades, and there is probably a direct correlation between gene expression level and grade progression. The increase of occludin expression level of tumor samples in grade I and II was significant ($P = 0.0365$ and $P = 0.0005$, respectively) in comparison with that of the normal sample.

With respect to the stage of disease and claudin and occludin expression levels, it was figured out that the majority of tumor samples had an increased expression level compared with the normal sample. According to Figure 4, the most average claudin expression level is respectively observed in stages II, III, and IV, while the expression level in stage II is about 3 to 4 times greater than the stages III and IV. There is probably no direct correlation between the stage progression and gene expression level. It must be mentioned that the number of specimens in stages I, II, and III was very low. The increase of claudin expression level was not significant in any of the stages ($P > 0.05$).

According to Figure 5, the average of occludin expression level in stages II, III, and IV has a remarkable increase; however, there is probably no direct relation between the stage progression and occludin expression level. It should be considered that the number of samples was very low in the stages I, II, and III. The increase of occludin expression level compared with the normal sample was significant only at stage III ($P = 0.0274$).

With regard to metastasis and claudin and occludin expression level, the tumor expression level in metastatic samples is more than the normal tissue. According to Figures 4 and 5, the highest average of expression level in both genes is observed in non-metastatic samples. It is worth considering that the number of metastatic samples was very low. The increased expression level of any of the genes was not significant in any of the cases ($P > 0.05$).

Examination of the tumor size and claudin expression level reveals that the highest average of gene expression level is observed in tumors less than 3 cm (Figure. 4). The mentioned level was about 3 and 8 times higher than that of the other groups with tumors between 3 and 5 cm and tumors larger than 5 cm, respectively. The increase of claudin expression level was significant in tumors larger than 5 cm ($P = 0.0322$).

According to Figure 5, which addresses the tumor size and occludin expression level, the higher expression level is observed for tumors within the size range of 3 to 5 cm. The observed expression level was about 5 and 10 times higher than the other categories. There is probably no direct relation between occludin expression level and the tumor size. The increase of occludin expression level in tumors smaller than 3 cm and larger than 5cm ($P = 0.0057$ and $P = 0.0325$, respectively) was significant in comparison with the normal sample.

**Discussion**

Epithelial mesenchymal transition (EMT) is a phenomenon which begins with changes in the tight junction and cytoskeleton and leads to extracellular matrix transformation and at last cell migration [11]. EMT is a highly protected cell program that controls a process in which differentiated polarized immobile epithelial cells can be transformed into mobile mesenchymal cells [12]. The mentioned process is vital for most fetal developmental events and can be re-enacted in a variety of diseases including cancer. Epithelial cell transformation and differentiation are the markers of tumor progression during invasion and metastasis phase. A number of important epithelial proteins such as E-cadherin, occludin, claudins, cytokeratins and catenin have been downregulated in cancerous cells with metastatic tendency [13]. Claudins, occludin and ZO (Zonula Occludens) proteins are likely to be specific obstacles for endothelium to play its functional role. These are the main proteins identified in tight junction membrane [14].

As suggested in previous studies, Fibroblast growth factor-10 (FGF-10), which plays a crucial role in EMT type I during the gastrulation, is dysregulated in cancer and can be considered to be involved in invasion and cell migration [15].

In the present study, claudin and occludin expression level has increased in majority of tumor tissues considering the grade and stage of disease. Their higher expression level may have prognostic significance in patients with colorectal cancer. However, it should be reminded that little change was noted in some of the samples due to performing the *in vivo* study and applying mechanisms that control these two genes. In a similar study addressing breast cancer, increasing the expression level of occludin gene was considered to be probably one of the effective factors in tissue transformation toward cancerous phase [16]. One of the studies examining the expression of claudin-5, claudin-7, and occludin in oral squamous cell carcinoma showed that claudin-7 expression can be effective in predicting OSCC and prognosis of the patients [17]. Moreover, another study on occludin expression in brain tumors suggested that the expression of occludin is highly related to the development of peritumoral brain edema in brain tumors and can be regarded as a factor for prognosis [18]. Furthermore, in a recent study focusing on claudin-1 expression in cervical cancer, it was proposed that claudin-1 is probably a molecular marker in squamous cervical cancer and potentially can be a diagnostic, prognostic, and therapeutic marker [19].

The tumor requires undergoing further metabolisms to increase its size; hence, it has more angiogenesis than a smaller tumor. Therefore, examination of the relation between gene expression and tumor size is of great value. Regarding the tumor size in the present study, the increase of the expression level of both claudin and occludin genes was significant in tumors larger than 5 cm. In another pertinent study, the protein expression of claudin 1, 3, 4, 5 and 7 and the clinical significance and association with tumor growth pattern in colon carcinoma were analyzed. The observed results suggested that claudin expression in colon carcinoma cells may have a progressive effect in colon carcinoma development and can be considered as a tumor marker [20]. In a recent study conducted on tumor malignancy in colorectal carcinoma, an association was found between Vimentin expression and tumor size; the expression decreased in larger tumor sizes. Additionally, increased expression of fibronectin was correlated with high tumor stages [21].

In the present research, the number of metastatic specimens was insufficient to make a reasonable conclusion; however, another study addressing claudin expression profiles of hepatocellular carcinoma and metastatic colorectal and pancreatic carcinomas revealed that claudin protein was highly expressed in liver metastases of colorectal adenocarcinoma. Moreover, it was indicated that distinct claudin expression profiles might provide better understanding of the pathobiology of these lesions and might be used for differential diagnosis [22].

**Conclusion**

Based on the results of this study, an increase in the expression level of claudin and occludin in colorectal cancer specimens was specified, which might be useful for the detection of colorectal cancer and its probable prognosis.

**Acknowledgements**

This study was supported by cancer research center (CRC) of Tehran University of Medical Sciences.